\documentclass[twocolumn,pra,aps,showpacs]{revtex4}

\usepackage{mathptmx}
\usepackage{subfigure}
\usepackage{psfrag,graphicx}
\usepackage{dcolumn}
\usepackage{amsmath,amssymb}
\usepackage{bm}
\usepackage{color}
\usepackage{latexsym}
\usepackage{epstopdf}
\usepackage{color}
\usepackage[english]{babel}
\usepackage{latexsym}
\usepackage{psfrag,graphicx}
\usepackage{subfigure}
\usepackage{amsmath}
\usepackage{amssymb}
\usepackage{amsfonts}
\usepackage{bm}
\usepackage{natbib}
\usepackage{epstopdf}
\usepackage{appendix}

\definecolor{mygrey}{gray}{0.35}
\definecolor{myblue}{rgb}{0.2,0.2,0.8}
\definecolor{myzard}{cmyk}{0,0,0.05,0}
\definecolor{mywhite}{rgb}{1,1,1}
\definecolor{mywhite}{rgb}{1,1,1}
\definecolor{myred}{rgb}{1,0.,0.3}

\usepackage[colorlinks=true,citecolor=myblue,linkcolor=myred]{hyperref}

\def\ba{\begin{align}}
\def\enda{\end{align}}
\def\bi{\begin{itemize}}
\def\ei{\end{itemize}}

\def\be{\begin{equation}}
\def\ee{\end{equation}}
\def\bea{\begin{eqnarray}}
\def\eea{\end{eqnarray}}
\def\bse{\begin{subequations}}
\def\ese{\end{subequations}}

\begin{document}

\title{Collective Modes in the Cooperative Jahn-Teller Model: Path Integral Approach}

\pacs{64.70.Tg, 03.67.Ac, 37.10.Ty, 71.70.Ej}

\author{Peter A. Ivanov }

\email{pivanov@phys.uni-sofia.bg}

\affiliation{Department of Physics, St. Kliment Ohridski University of Sofia, James
Bourchier 5 Boulevard, 1164 Sofia, Bulgaria}

\begin{abstract}
We discuss analytical approximations to the ground-state phase diagram and the elementary excitations of the cooperative Jahn-Teller model describing strongly correlated spin-boson system on a lattice in various quantum optical systems. Based on the mean-field theory approach we show that the system exhibits quantum magnetic structural phase transition which leads to magnetic ordering of the spins and formation of the bosonic condensates. We determine existing of one gapless Goldstone mode and two gapped amplitude modes inside the symmetry-broken phase.
\end{abstract}

\maketitle

\section{Introduction}
Over the last few years there has been a great deal of interest in studying the many-body physics of strongly correlated spins-bosons lattice models using a table top experimental quantum-optical systems \cite{Georgescu2014}. A prominent example is the recent experimental demonstration of the quantum phase transition from Mott insulator state to superfluid state of spin-boson excitations in a system of trapped ions \cite{Toyoda2013}. It was shown that with the spontaneous breaking of continuous $U(1)$ symmetry at the quantum phase transition in such models, including for example Jaynes-Cummings-Hubbard system \cite{Schmidt2010} and Dicke-like system \cite{Baksic2014,Xiang2013}, two-types of collective excitations emerge. One is the the gapless Goldstone mode, and the other is the gapped amplitude mode corresponding to phase and density fluctuations. Such amplitude modulation of the order parameter is referred to as Higgs mode generated by a physical mechanism analogous as the Higgs boson in high energy physics. Recently, the amplitude mode was experimentally observed in a system of strongly interacting condensate of ultracold atoms near the superfluid-insulator phase transition \cite{Bissbort2011,Endres2012} opened fascinating prospect for exploring the condensed matter excitations under controlled conditions.

In this work we present study of the collective hybrid spin-boson excitations in the cooperative Jahn-Teller (cJT) model. Originally, the Jahn-Teller model was introduced to explain the distortions and the nondegenerate energy levels in molecules, via the strong interaction between the localized electronic states and the vibrations of the nuclei \cite{Englman,Bersuker}. In solids, the cJT effect leads to structural phase transition and  magnetic ordering of the spins. Furthermore, the collective effects induced by the Jahn-Teller coupling may explain the transition of some solids, such as fullerene compounds, to high-temperature superconductors \cite{Dunn2004}. With the current quantum optical technologies the cJT model can be realized in laser or magnetically driven ion crystal \cite{Porras2012,Ivanov2013} and cavity/circuit QED systems based on superconducting qubits in transmission line resonators \cite{Larson2008,Dereli}. Here we focus on $E\otimes e$ Jahn-Teller model which possess continuous $U(1)$ symmetry. We use path integral approach to describe analytically the quantum magnetic structural phase transition with the formation of bosonic condensates and magnetic ordering of the spins in the cJT model. Within the framework of the saddle-point approximation we determine the mean-field solution and then consider the quantum fluctuations around the mean-field result. We show that the energy spectrum of the cJT system consists of three collective excitations branches. In the symmetry broken phase we find a linear gapless Goldstone mode and two gapped amplitude modes.

The paper is organized as follows. In Sec. \ref{CJTm} we introduce the cJT model and consider the associated continuous $U(1)$ symmetry. In Sec. \ref{implementation} we discuss a possible scheme for the experimental realization of the cJT model using quantum optical systems. In Sec. \ref{PIF} we turn to the path integral treatment of the model and determine its saddle-point. In Sec. \ref{QF} we consider the quantum fluctuations around the mean-field solution and find the elementary excitations of the cJT model. In Sec. \ref{MP} we discuss many-body spectroscopy protocol to detect the quantum phases and the collective excitations. Finally, the conclusions are presented in Sec. \ref{C}.



\section{Cooperative Jahn-Teller Model}\label{CJTm}

We consider a chain of $N$ spins with states $\left|\uparrow_{j}\right\rangle$, $\left|\downarrow_{j}\right\rangle$ each one coupled symmetrically with two boson species ($\epsilon=x,y$ and $\hbar=1$ from now on),
\begin{eqnarray}
&&\hat{H}_{\rm cJT}=\hat{H}_{\rm s}+\hat{H}_{\rm t-b}+\hat{H}_{\rm I},\quad \hat{H}_{\rm s}=\frac{\omega_{z}}{2}\sum_{j=1}^{N}\sigma_{j}^{z}, \notag\\
&&\hat{H}_{\rm t-b}=\sum_{\epsilon}\sum_{j=1}^{N}\Delta_{j}\hat{a}_{\epsilon,j}^{\dag}\hat{a}_{\epsilon,j}+\sum_{\epsilon}\sum_{j>l}^{N}t_{j,l}(\hat{a}_{\epsilon,j}^{\dag}\hat{a}_{\epsilon,l}+{\rm H.c}),\notag\\
&&\hat{H}_{\rm I}=\frac{g}{\sqrt{2}}\sum_{j=1}^{N}\{\sigma_{j}^{x}(\hat{a}_{x,j}^{\dag}+\hat{a}_{x,j})+\sigma_{j}^{y}(\hat{a}_{y,j}^{\dag}+\hat{a}_{y,j})\}.\label{HcJT}
\end{eqnarray}
The term $\hat{H}_{\rm s}$ describes the energy of the effective spins with frequency $\omega_{z}$, where $\sigma_{j}^{c}$ ($c=x,y,z$) are the Pauli matrices for the spin at site $j$. Note that this term also can represent the coupling with applied external magnetic field. The tight-binding term $\hat{H}_{\rm t-b}$ describes the delocalization of the two bosonic species $\epsilon=x,y$ between different lattice sites with hopping matrix elements $t_{j,l}<0$ and on-site boson energy $\Delta_{j}$, where $\hat{a}_{\epsilon,j}^{\dag}$, $\hat{a}_{\epsilon,j}$ are the respective creation and annihilation operators of boson at site $j$. The last term in (\ref{HcJT}) describes the $E\otimes e$ symmetrical Jahn-Teller interaction between the spins and two boson species with coupling strength $g$.

Alternatively, the spin-boson interaction can be expressed in terms of right $\hat{a}_{{\rm r},j}^{\dag}=(\hat{a}_{x,j}^{\dag}+{\rm i}\hat{a}_{y,j}^{\dag})/\sqrt{2}$ and left $\hat{a}_{{\rm l},j}^{\dag}=(\hat{a}_{x,j}^{\dag}-{\rm i}\hat{a}_{y,j}^{\dag})/\sqrt{2}$ chiral operators, which yield
\begin{equation}
\hat{H}_{\rm I}=g\sum_{j=1}^{N}\{\sigma_{j}^{+}(\hat{a}_{{\rm r},j}+\hat{a}_{{\rm l},j}^{\dag})+\sigma_{j}^{-}(\hat{a}_{{\rm r},j}^{\dag}+\hat{a}_{{\rm l},j})\},\label{HI}
\end{equation}
where $\sigma_{j}^{\pm}$ are the corresponding spin raising and lowering operators. We note that because of the contra-rotating terms $\sigma_{j}^{+}\hat{a}_{{\rm l},j}^{\dag}$ and $\sigma_{j}^{-}\hat{a}_{{\rm r},j}$ in (\ref{HI}) the total number of spin and boson excitations is not conserved. Instead of that, the Hamiltonian $H_{\rm cJT}$ commute with the operator $\hat{C}=\sum_{j=1}^{N}(\hat{a}_{{\rm r},j}^{\dag}\hat{a}_{{\rm r},j}-\hat{a}_{{\rm l},j}^{\dag}\hat{a}_{{\rm l},j}+\sigma_{j}^{z}/2)$. The latter implies that the cJT Hamiltonian (\ref{HcJT}) possesses continuous $U(1)$ symmetry implemented by the action of the operator $\hat{R}(\phi)=e^{{\rm i}\phi \hat{C}}$, which gives
\begin{eqnarray}
&&\hat{R}(\phi)\hat{a}_{{\rm r},j}^{\dag}\hat{R}(\phi)^{\dag}=e^{{\rm i}\phi}\hat{a}_{{\rm r},j}^{\dag},\quad \hat{R}(\phi)\hat{a}_{{\rm l},j}^{\dag}\hat{R}(\phi)^{\dag}=e^{-{\rm i}\phi}\hat{a}_{{\rm l},j}^{\dag},\notag\\
&&\hat{R}(\phi)\sigma_{j}^{\pm}\hat{R}(\phi)^{\dag}=e^{\pm {\rm i}\phi}\sigma_{j}^{\pm},\label{symmetry}
\end{eqnarray}
such that we have $\hat{R}(\phi)\hat{H}_{{\rm cJT}}\hat{R}(\phi)^{\dag}=\hat{H}_{{\rm cJT}}$. Finally, it is convenient to work in representation, where the tight-binding term $\hat{H}_{\rm t-b}$ in (\ref{HcJT}) is diagonal. Indeed, performing the transformation in to the momentum space $\hat{a}_{\gamma,k}=\sum_{j=1}^{N}b_{k,j}\hat{a}_{\gamma,j}$, ($\gamma={\rm r,l}$ from now on), where $b_{k,j}$ are the normal mode wave functions yield $\hat{H}_{\rm t-b}=\sum_{\gamma}\sum_{k=1}^{N}\Delta_{k}\hat{a}_{\gamma,k}^{\dag}\hat{a}_{\gamma,k}$ with $\Delta_{k}$ being the collective mode energies. Hereafter we assume that $\Delta_{k}>0$ with minimum at $k=0$ corresponding to the center-of-mass mode.

In the following we discuss the realization of the cJT model using quantum optical systems.

\section{Implementation with quantum optical systems}\label{implementation}
The cJT model comprises of two bosonic species and spin degrees of freedom, the specific interpretation of which depends on the actual physical system. One possible experimental setup for the realization of the model is based on the laser cooled trapped ions \cite{Wineland,Schneider2012}. In that case the bosonic degrees of freedom represent the local phonons, which quantify the small radial ion oscillations around the equilibrium positions \cite{Ivanov2009,Porras2004}. The hopping term $\hat{H}_{\rm t-b}$ in (\ref{HcJT}) describes the Coulomb-mediated long-range phonon hopping dynamics with hopping elements $t_{j,l}$ and on-site frequency $\Delta_{j}$. The spins are implemented by the internal two metastable levels of the trapped ions where $\omega_{z}$ is the effective spin frequency. The desired Jahn-Teller coupling can be realized by the interaction of the spins with an oscillating magnetic field gradient \cite{Porras2012,Ivanov2013}. Alternative realization of the Jahn-Teller coupling is based on the interaction of the ions with laser beams propagating in two orthogonal directions tuned near the respective red and blue sidebands \cite{Ivanov2014}. The ion trap based realization of the cJT model offers unique opportunity to easy tuning the parametric regime of the couplings by adjusting for example the trap frequencies and the laser intensity. Although with the current ion technologies the realization of the model is restricted to one-dimension where the ions are placed in a chain, considerable progress is achieved to scaling to two-dimensional ion trap network where the ions are trapped in individual potential wells \cite{Sterling2014,Wilson2014}.

On the other hand lattice spin-boson models can be realized naturally in cavity and circuit QED systems \cite{Greentree2006,Angelakis2007,Hartmann2007}. Here the bosons represent a single or several quantized modes inside of an electromagnetic resonators, while the spin degrees of freedom are implemented either by real two-level atom, or artificial atoms such as quantum dot or superconducting circuit. Recently it was shown that the interaction between the quantum dot \cite{Larson2008} or superconducting circuit \cite{Dereli} with two cavity modes leads to the Jahn-Teller coupling. Arranging atoms and resonators in the form of lattice can realize our model (\ref{HcJT}), with coupling between the resonators provided by the photon hopping.

In the next section, we turn to the field-theoretical treatment of the cJT model. This method allows us to determine the stationary saddle-point of the model and then consider the small fluctuations around the mean-field result.

\section{Functional Integral Representation of the cooperative Jahn-Teller model}\label{PIF}
\subsection{Path integral approach to the cJT model}
In the functional integral treatment, the second quantized Hamiltonian of the model is translated to the phase representation with the help of the path integral formalism. In this approach the boson operators are replaced with their associated fields, namely $\hat{a}_{\gamma,k}\rightarrow \alpha_{\gamma,k}(\tau)$, $\hat{a}_{\gamma,k}^{\dag}\rightarrow \alpha_{\gamma,k}^{*}(\tau)$ where $\tau$ is the imaginary time \cite{Auerbach}. For the spin-degrees of freedom we choose a spin-coherent representation with coherent-state parameterized by the independent polar $\theta_{j}$ and azimuthal $\varphi_{j}$ angles, respectively,
\begin{equation}
|n_{j}\rangle=\cos\left(\frac{\theta_{j}}{2}\right)\left|\uparrow_{j}\right\rangle+e^{{\rm i}\varphi_{j}}\sin\left(\frac{\theta_{j}}{2}\right)\left|\downarrow_{j}\right\rangle.\label{nvector}
\end{equation}
The spin operators are replaced by the corresponding Bloch vector $\vec{n}_{j}=[n_{x,j},n_{y,j},n_{z,j}]$ whose components are the expectation values of the Pauli matrices with respect to the state (\ref{nvector}) which gives
\begin{equation}
\vec{n}_{j}=[\sin(\theta_{j})\cos(\varphi_{j}),\sin(\theta_{j})\sin(\varphi_{j}),\cos(\theta_{j})].
\end{equation}
The Bloch vector has a unit length $\vec{n}_{j}^{2}=1$ and specifies the orientation of spin at site $j$. Having this in hand the partition function for the cJT model can be expressed as
\begin{equation}
Z(\beta)=\int\prod_{\gamma,k}\prod_{j}D\alpha_{\gamma, k}^{*}(\tau)D\alpha_{\gamma, k}(\tau)D\vec{n}_{j}(\tau)
\delta(\vec{n}_{j}^{2}-1)e^{-S},
\end{equation}
with the Euclidian action given by
\begin{eqnarray}
S&=&\int_{0}^{\beta}d\tau\{\sum_{\gamma,k}(\alpha_{\gamma,k}^{*}\frac{\partial\alpha_{\gamma,k}}{\partial\tau}+\Delta_{k}\alpha_{\gamma,k}^{*}\alpha_{\gamma,k})
+\frac{\omega_{z}}{2}\sum_{j}\cos(\theta_{j})\notag\\
&&+\frac{g}{2}\sum_{j,k}\sin(\theta_{j})\{e^{{\rm i}\varphi_{j}}(b_{k,j}^{*}\alpha_{{\rm r},k}+b_{k,j}\alpha_{{\rm l},k}^{*})+{\rm c.c}\}\}+S_{\rm B},\label{S}
\end{eqnarray}
where $\beta=1/T$ is the inverse temperature.
The Berry phase contribution to the action (\ref{S}) from the spin-degrees of freedom is given by
\begin{equation}
S_{\rm B}=\sum_{j}\int_{0}^{\beta}\langle n_{j}|\frac{\partial}{\partial\tau}|n_{j}\rangle={\rm i}\sum_{j}\int_{0}^{\beta}d\tau
\sin^{2}\left(\frac{\theta_{j}}{2}\right)\frac{\partial\varphi_{j}}{\partial\tau}.\label{berry}
\end{equation}
Finally, we require that the corresponding bosonic fields have periodic boundary conditions $\alpha_{\gamma,k}(\beta)=\alpha_{\gamma,k}(0)$ and $\alpha_{\gamma,k}^{*}(\beta)=\alpha_{\gamma,k}^{*}(0)$. The same condition is hold and for the spin variables, where we have $\theta_{j}(\beta)=\theta_{j}(0)$ and $\varphi_{j}(\beta)=\varphi_{j}(0)$.

\subsection{Saddle-Point Approximation}
Next we consider the classical equation of motion, which are determined by the condition, that the variation of the action (\ref{S}) with respect to the field variables should vanish,
\begin{equation}
\frac{\delta S}{\delta \theta_{j} }=0, \quad \frac{\delta S}{\delta \varphi_{j} }=0, \quad \frac{\delta S}{\delta \alpha_{\gamma,k}^{*}}=0.\label{Var}
\end{equation}
Note that the same condition is also satisfied for the bosonic field $\alpha_{\gamma,k}$. The term classical refers to a mean-field solution, i.e., disregarding the quantum fluctuations. The variation of the action $S$ with respect of the spin-degrees of freedom gives the following equations of motion
\begin{eqnarray}
{\rm i}\sin(\theta_{j})\frac{\partial\varphi_{j}}{\partial\tau}&=&\omega_{z}\sin(\theta_{j})-g\cos(\theta_{j})
\sum_{k}\{e^{{\rm i}\varphi_{j}}(b_{k,j}^{*}\alpha_{{\rm r},k} \notag\\
&&+b_{k,j}\alpha_{{\rm l},k}^{*})+{\rm c.c}\},\notag\\
\frac{\partial\theta_{j}}{\partial\tau}&=&g\sum_{k}\{e^{{\rm i}\varphi_{j}}(b_{k,j}^{*}\alpha_{{\rm r},k}
+b_{k,j}\alpha_{{\rm l},k}^{*})-{\rm c.c}\})\label{spin_eq},
\end{eqnarray}
where the dynamics follow from the Berry phase term (\ref{berry}). The third condition in (\ref{Var}) reads
\begin{eqnarray}
\frac{\partial\alpha_{{\rm r},k}}{\partial\tau}&=&-\Delta_{k}\alpha_{{\rm r},k}-\frac{g}{2}\sum_{j}b_{k,j}(n_{x,j}-{\rm i}n_{y,j}),\notag\\
\frac{\partial\alpha_{{\rm l},k}}{\partial\tau}&=&-\Delta_{k}\alpha_{{\rm l},k}-\frac{g}{2}\sum_{j}b_{k,j}(n_{x,j}+{\rm i}n_{y,j}).\label{alpha}
\end{eqnarray}
We note that because the action $S$ (\ref{S}) is invariant with respect to $U(1)$ transformation specified in Eq. (\ref{symmetry}), the corresponding equations of motion (\ref{spin_eq}) and (\ref{alpha}) obey the same symmetry. Although the Eq. (\ref{spin_eq}) describes the dynamics of the spin-degree of freedom it is not expressed in terms of $\vec{n}_{j}$. One way to remedy this is to introduce a new set of two orthogonal to $\vec{n}_{j}$ vectors, namely $\vec{\theta}_{j}=[\cos(\theta_{j})\cos(\varphi_{j}),\cos(\theta_{j})\sin(\varphi_{j}),-\sin(\theta_{j})]$ and $\vec{\varphi}_{j}=[-\sin(\varphi_{j}),\cos(\varphi_{j}),0]$ which form an orthogonal triad $\vec{\theta}_{j}\times\vec{\varphi}_{j}=\vec{n}_{j}$. Then the Eq. (\ref{spin_eq}) is rewritten as follows
\begin{eqnarray}
\frac{{\rm i}}{2}\vec{\varphi}_{j}\cdot\frac{\partial \vec{n}_{j}}{\partial\tau}&=&\frac{\omega_{z}}{2}(\vec{\varphi}_{j}+\vec{\theta}_{j})\cdot\frac{\partial\vec{n}_{j}}{\partial\varphi_{j}}-g\vec{\theta}_{j}\cdot\vec{\alpha}_{j},\notag\\
\frac{{\rm i}}{2}\vec{\theta}_{j}\cdot\frac{\partial\vec{n}_{j}}{\partial\tau}&=&g\vec{\varphi}_{j}\cdot\vec{\alpha}_{j}.\label{spin1}
\end{eqnarray}
Here we have introduce the vector notation $\vec{\alpha}_{j}=\sqrt{2}[\Re{\alpha_{x,j}},\Re{\alpha_{y,j}},0]$ for the two bosonic fields. Finally, one can combine the two equations in (\ref{spin1}) in a vector form which yield
\begin{equation}
\frac{{\rm i}}{2}\frac{\partial\vec{n}_{j}}{\partial\tau}=\frac{\omega_{z}}{2}\vec{B}\times\vec{n}_{j}+g\vec{\alpha}_{j}\times\vec{n}_{j},\label{spin_vector}
\end{equation}
where we use $\frac{\partial\vec{n}_{j}}{\partial\varphi_{j}}=\vec{B}\times \vec{n}_{j}$. Now we are in position to interpret the dynamical equation for the spins. The first term in (\ref{spin_vector}) represent the effect of the externally applied magnetic field along the $z$ direction, $\vec{B}=[0,0,1]$, such that for $g=0$ the spins will perform precession with frequency determined by $\omega_{z}$. On the other hand the spin-boson interaction gives rise to an effective magnetic field $\vec{\alpha}_{j}$ experienced by the spin at lattice site $j$. In the case when the bosonic fields describe the motional degrees of freedom of the spins in two orthogonal directions, the effective magnetic field $\vec{\alpha}_{j}$ becomes position-dependent, which is in close analogy with the Rashba spin-orbit coupling in the quantum spin Hall effect \cite{Sinova2004}. Because the vectors $\vec{B}$ and $\vec{\alpha}_{j}$ are always orthogonal, the spins execute precession along the axis $45^{0}$ to both magnetic fields. We note that although Eq. (\ref{spin_vector}) is purely local in a sense that it only depends on the lattice index $j$, the components of $\vec{\alpha}_{j}$ depend on the boson fields at different sites due to the tunneling elements $t_{j,l}$. As we will see below such a tight-binding lattice dynamics of the two bosonic species strongly coupled to the spins are capable of forming magnetic ordering and bosonic condensates.

The stationary saddle-point is obtained by the solution of Eq. (\ref{Var}) with the requirement that $\alpha_{\gamma,k}(\tau)=\bar{\alpha}_{\gamma,k}$ and $\vec{n}_{j}(\tau)=\vec{\bar{n}}_{j}$. Then we derive the following set of algebraic equations for the spin-degrees of freedom
\begin{eqnarray}
&&\omega_{z}\sin(\bar{\theta}_{j})=-\cos(\bar{\theta}_{j})\sum_{l}J_{j,l}\sin(\bar{\theta}_{l})\cos(\bar{\varphi}_{j}-\bar{\varphi}_{l}),\notag\\
&&\sum_{l}J_{j,l}\sin(\bar{\theta}_{l})\sin(\bar{\varphi}_{j}-\bar{\varphi}_{l})=0,\label{mf_s}
\end{eqnarray}
where $J_{j,l}=2\sum_{k}\frac{g^{2}}{\Delta_{k}}\Re\{b_{k,j}b_{k,l}^{*}\}$ and respectively for the bosonic fields
\begin{eqnarray}
&&\bar{\alpha}_{{\rm r},k}=-\frac{g}{2\Delta_{k}}\sum_{j}b_{k,j}\sin(\bar{\theta}_{j})e^{-{\rm i}\bar{\varphi}_{j}},\notag\\
&&\bar{\alpha}_{{\rm l},k}=-\frac{g}{2\Delta_{k}}\sum_{j}b_{k,j}\sin(\bar{\theta}_{j})e^{{\rm i}\bar{\varphi}_{j}}\label{mf_b}.
\end{eqnarray}
Apparently, the system (\ref{mf_s}) has a trivial solution $\sin(\bar{\theta}_{j})=0$ for ($j=1,2,\ldots,N$) which implies $\bar{\alpha}_{\gamma,k}=0$. Assuming $\sin(\bar{\theta}_{j})\neq 0$, the condition $\bar{\varphi}_{j}=\bar{\varphi}_{l}=\bar{\varphi}$ solved the second equation in (\ref{mf_s}). The latter is the arbitrary choice for a direction of spontaneous symmetry breaking where the system chooses a direction along which to order. Hear after we assume $\bar{\varphi}=0$, such that the Bloch vector becomes $\vec{\bar{n}}_{j}=[\sin(\bar{\theta}_{j}),0,\cos(\bar{\theta}_{j})]$ indicating that the spins are aligned in the $xz$ plane. Let us now discuss the homogenous limit $\bar{\theta}_{j}=\bar{\theta}$ neglecting any boundary effect. In this limit Eqs. (\ref{mf_s}) and (\ref{mf_b}) can be solved exactly, which yield
\begin{eqnarray}
&&\cos(\bar{\theta})=-1,\quad \bar{\alpha}_{\gamma,k}=0,\quad g< g_{{\rm c}},\notag\\
&&\cos(\bar{\theta})=-\frac{g_{{\rm c}}^{2}}{g^{2}},\quad \bar{\alpha}_{\gamma,k}=-\frac{g\sqrt{N}}{2\Delta_{k}}\sin(\bar{\theta})\delta_{k,0},\quad
g>g_{\rm {c}}\label{sp}
\end{eqnarray}
The solution (\ref{sp}) corresponds to the classical ground-state of the cJT model. For a coupling smaller than the critical value of $g_{{\rm c}}=\sqrt{\Delta_{0}\omega_{z}/2}$ ($g<g_{\rm c}$) the system is in a normal state where the Bloch vector for each spin points along the $-z$ direction and $\bar{\alpha}_{\gamma,k}=0$. Increasing the coupling through $g_{{\rm c}}$ ($g>g_{\rm c}$) drives the system to undergo a quantum phase transition to a ferromagnetic ordering of spins in $xz$ plane and condensation of the two boson species in the lowest energy mode $k=0$.
Here we emphasize that for a linear ion crystal with positive hopping amplitude the saddle-point approximation is not applied straightforward. This problem can be overcome by applying a canonical transformation to the operators $\hat{a}_{\epsilon,j}\rightarrow (-1)^{j}\hat{a}_{\epsilon,j}$ and $\sigma_{j}^{\epsilon}\rightarrow (-1)^{j}\sigma_{j}^{\epsilon}$ in (\ref{HcJT}) \cite{Porras2012,Mering2009}. After this transformation to the staggered spin-boson basis the cJT model (\ref{HcJT}) is unchanged, but the tunneling is modified to $t_{j,l}^{\rm stagg}=(-1)^{j-l}t_{j,l}$. The ferromagnetic spin order in the new basis, corresponds to an antiferromagnetic order in the physical basis, in which ions alternate spin direction and position.

In the following section we study the low-energy spectra of the cJT model in terms of collective excitations. We expand the action of the system around its saddle-point up to second order in the spin and bosonic fields. This leads to a Gaussian integral which can be evaluated.

\section{Quantum Fluctuations around the saddle-point}\label{QF}
\subsection{Linear parametrization}
Having described the saddle-point solution, we now consider the low-energy excitations of the cJT model in the symmetry-broken phase. For the bosonic fields we can use the standard linear parametrization
\begin{equation}
\alpha_{\gamma,k}=\bar{\alpha}_{\gamma,k}+\delta\alpha_{\gamma,k},
\end{equation}
where $\delta\alpha_{\gamma,k}$ describes the quantum fluctuations around the order parameter $\bar{\alpha}_{\gamma,k}$. In order to account the spin fluctuations around the state $\vec{\bar{n}}_{j}$ for each spin at site $j$ we first perform rotation of the Bloch vector $\vec{n}_{j}=\hat{R}(\bar{\theta})\vec{n}_{j}^{\prime}$ with rotation matrix given by
\begin{figure}[h]
\includegraphics[width=0.45\textwidth]{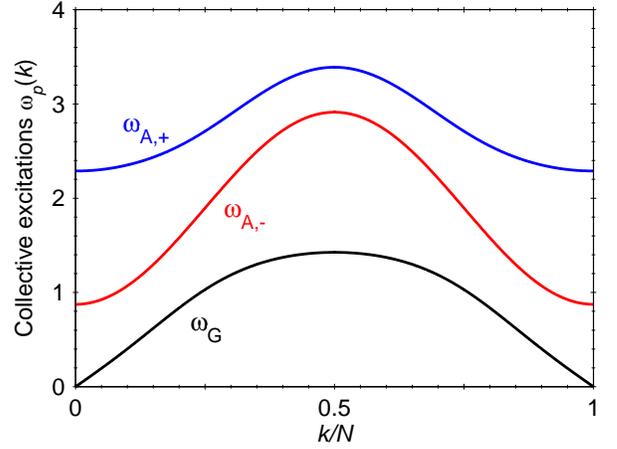}
\caption{(Color online) The dispersions of the three branch frequencies versus $k$. We set $\Delta/g=1$, $t/g=0.5$, and $\omega_{z}/g=1$. We have assumed position independent bosonic frequency $\Delta_{j}=\Delta+2t$.}
\label{fig1}
\end{figure}
\begin{equation}
\hat{R}(\bar{\theta}) =\left[
\begin{array}{ccc}
\cos(\bar{\theta}) & 0 & \sin(\bar{\theta})\\
0 & 1 & 0 \\
-\sin(\bar{\theta}) & 0 & \cos(\bar{\theta})%
\end{array}%
\right].\label{R}
\end{equation}%
Note that transformation of the Bloch vector implies rotation of the spin-coherent state specified by $|n_{j}^{\prime}\rangle=e^{{\rm i}\frac{\bar{\theta}}{2}\sigma_{j}^{y}}|n_{j}\rangle$. The rotation matrix $\hat{R}(\bar{\theta})$ is determined in a such a way that transform $\vec{\bar{n}}_{j}$ to a new reference Bloch vector $\vec{\bar{n}}_{j}^{\prime}=[0,0,1]$ which points along the $z$ direction.
Assuming that the spin and bosonic fluctuations are small in a sense that one can keep only the quadratic terms, such that $n_{z,j}^{\prime}=\sqrt{1-n_{x,j}^{\prime 2}-n_{y,j}^{\prime2}}\approx 1-(n_{x,j}^{\prime 2}+n_{y,j}^{\prime 2})/2$ we obtain
\begin{eqnarray}
S&=&S_{\rm B}^{\prime}+\int_{0}^{\beta}d\tau\{\sum_{\epsilon,k}(\delta\alpha_{\epsilon,k}^{*}\frac{\partial\delta\alpha_{\epsilon,k}}{\partial\tau}+\Delta_{k}\delta\alpha_{\epsilon,k}^{*}\delta\alpha_{\epsilon,k})\notag\\
&&+\frac{\Omega}{4}\sum_{j}(n_{x,j}^{\prime 2}+n_{y,j}^{\prime 2})+\frac{g}{\sqrt{2}}\sum_{j,k}n_{y,j}^{\prime}
\{b_{k,j}^{*}\delta\alpha_{y,k}+b_{k,j}\delta\alpha_{y,k}^{*}\}
\notag\\
&&+\frac{g}{\sqrt{2}}\cos(\bar{\theta})\sum_{j,k}n_{x,j}^{\prime}
\{b_{k,j}^{*}\delta\alpha_{x,k}+b_{k,j}\delta\alpha_{x,k}^{*}
\}\},\label{S_f}
\end{eqnarray}
where the linear terms in the field fluctuations vanishes due to the conditions Eqs. (\ref{mf_s}) and (\ref{mf_b}) for $g>g_{\rm c}$. Here  $\Omega=\omega_{z}/|\cos(\bar{\theta})|$ is the renormalised spin frequency and $S_{\rm B}^{\prime}$ is the Berry phase term in the rotating basis. Note that the $S_{\rm B}^{\prime}$ is invariant with respect to the rotation transformation, which implies that $S_{{\rm B}}^{\prime}=\sum_{j}\int_{0}^{\beta}\langle n_{j}^{\prime}|\frac{\partial}{\partial\tau}|n_{j}^{\prime}\rangle$. Up to quadratic terms in the spin fluctuation fields the Berry phase can be written as
\begin{equation}
S_{\rm B}^{\prime}=\frac{{\rm i}}{4}\sum_{j}\int_{0}^{\beta}d\tau\left(n_{x,j}^{\prime}\frac{\partial n_{y,j}^{\prime}}{\partial\tau}-n_{y,j}^{\prime}\frac{\partial n_{x,j}^{\prime}}{\partial\tau}\right).
\end{equation}
We emphasize that in order to describe the collective excitations around the mean-field solution one needs to identify the number of conjugate pairs. In our model presented here the bosonic fields $\delta\alpha_{\epsilon,k}$ and $\delta\alpha_{\epsilon,k}^{*}$ in (\ref{S_f}) are canonically conjugate variables, which leads to two independent degree of freedom. On the other hand the pairs $n_{+,j}^{\prime}=(n_{x,j}^{\prime}+{\rm i}n_{y,j}^{\prime})/2$ and $n_{-,j}^{\prime}=(n_{x,j}^{\prime}-{\rm i}n_{y,j}^{\prime})/2$ represent conjugate quantities corresponding to the spin fluctuations, which implies that one can expect in total three collective modes. In order to obtain the low-energy excitations, it is convenient to transform the spin fields in the momentum representation using $n_{+,j}^{\prime}=\sum_{k}b_{k,j}n_{+,k}^{\prime}$ and $n_{-,j}^{\prime}=\sum_{k}b_{k,j}^{*}n_{-,k}^{\prime}$, which yield
\begin{eqnarray}
S&=&\int_{0}^{\beta}\sum_{k}\{n_{-,k}^{\prime}\frac{\partial n_{+,k}^{\prime}}{\partial\tau} +\sum_{\epsilon}(\delta\alpha_{\epsilon,k}^{*}\frac{\partial\delta\alpha_{\epsilon,k}}{\partial\tau}+\Delta_{k}\delta\alpha_{\epsilon,k}^{*}\delta\alpha_{\epsilon,k})\notag\\
&&+\Omega n_{+,k}^{\prime}n_{-,k}^{\prime}-\frac{g}{\sqrt{2}}(n_{+,k}^{\prime}-n_{-,-k}^{\prime})(\delta\alpha_{y,-k}^{*}-\delta\alpha_{y,k})
\notag\\
&&+\frac{g}{\sqrt{2}}\cos(\bar{\theta})(n_{+,k}^{\prime}+n_{-,-k}^{\prime})(\delta\alpha_{x,-k}^{*}+\delta\alpha_{x,k}).\label{S_k}
\end{eqnarray}
The action (\ref{S_k}) is quadratic in the field fluctuations, which lead to Gaussian functional integral. To diagonalize (\ref{S_k}) one can introduce harmonic oscillator degrees of freedom for each pair of conjugate variables, such that we obtain (see, Appendix \ref{A})
\begin{eqnarray}
S&=&\int_{0}^{\beta}d\tau\sum_{k}\{{\rm i}\sum_{a=1}^{3}\frac{\partial p_{a,k}}{\partial\tau}q_{a,k}+\frac{1}{2}\sum_{a=1}^{3}p_{a,k}p_{a,-k}\notag\\
&&+\frac{1}{2}\sum_{a,a^{\prime}=1}^{3}B_{a,a^{\prime}}^{(k)}q_{a,k}q_{a^{\prime},-k}\},\label{S_final}
\end{eqnarray}
where the coupling matrix $B_{a,a^{\prime}}^{(k)}$ is given by
\begin{equation}
B_{a,a^{\prime}}^{(k)} =\left[
\begin{array}{ccc}
\Delta_{k}^{2} & -g_{\rm c}^{2}\sqrt{\frac{2\Delta_{k}}{\Delta_{0}m_{+}(k)}} & -g_{\rm c}^{2}\sqrt{\frac{2\Delta_{k}}{\Delta_{0}m_{-}(k)}}\\
-g_{\rm c}^{2}\sqrt{\frac{2\Delta_{k}}{\Delta_{0}m_{+}(k)}} & \frac{\varepsilon_{k}^{2}}{m_{+}(k)} & \frac{\varepsilon_{k}^{2}-\Delta_{k}^{2}}{\sqrt{m_{+}(k)m_{-}(k)}} \\
-g_{\rm c}^{2}\sqrt{\frac{2\Delta_{k}}{\Delta_{0}m_{-}(k)}} & \frac{\varepsilon_{k}^{2}-\Delta_{k}^{2}}{\sqrt{m_{+}(k)m_{-}(k)}} & \frac{\varepsilon_{k}^{2}}{m_{-}(k)}%
\end{array}%
\right],\label{B_matrix}
\end{equation}%
with $\varepsilon_{k}^{2}=(\Delta_{k}^{2}+\Omega^{2})/2$ and $m_{\pm}(k)=(1\pm\sqrt{\Delta_{0}/\Delta_{k}})^{-1}$. The dispersion relation of field fluctuations around the ground-state configuration, i.e., the collective spin-boson excitations, can be found by solving the eigenvalue problem $\sum_{a}B_{a,a^{\prime}}^{(k)}u_{a}^{(p)}(k)=\omega_{p}^{2}(k)u_{a^{\prime}}^{(p)}(k)$ with $p=1,2,3$. The result is summarized in Fig. \ref{fig1} where are shown the three branches of collective excitations, assuming periodic boundary conditions with nearest-neighbours bosonic tunneling $t_{j,l}=-t(\delta_{j,l+1}+\delta_{j,l-1})$ and bosonic dispersion $\Delta_{k}=\Delta+2t\{1-\cos(2\pi k/N)\}$ \cite{Nevado2013}. The lowest-lying branch correspond to the gapless Goldstone mode $\omega_{\rm G}$, which is linear for small $k$, i.e., $\omega_{\rm G}= c_{\rm s}2\pi k/N+O(k^{2})$ with characteristic slope $c_{\rm s}=2g^{2}\sin(\bar{\theta})\sqrt{t\Delta/(\Delta^{4}+4g^{4}\sin^{2}(\bar{\theta}))}$. The other two excitations, the so-called amplitude modes, remain gapped with $\omega_{{\rm A},\pm}=\Delta_{\pm}+O(k^{2})$, where the gaps are given by
\begin{equation}
\Delta_{\pm}^{2}=\frac{\Omega^{2}}{2}+\Delta^{2}\pm\sqrt{\frac{\Omega^{4}}{4}+4g_{\rm c}^{4}}.
\end{equation}
So far, we have discussed the quantum fluctuations of the bosonic fields $\alpha_{\gamma,k}$ around their classical configuration. Because, in the symmetry broken phase the saddle-point solution predicts formation of bosonic condensates, it is naturally to express the cJT action in terms of density and phase of the respective condensate. Such a treatment allows us to connect the density fluctuations and the local phase of the condensates with the creation of the energy gaps in the spectra of cJT model. A convenient way to do this is to adopt the polar parametrization of the bosonic fields.
\subsection{Polar decomposition}
Let us choose the nonlinear polar parametrization
\begin{equation}
\alpha_{\gamma,j}=\sqrt{\bar{\rho}+\delta\rho_{\gamma,j}}e^{{\rm i}\zeta{\gamma,j}},\label{polar}
\end{equation}
of the bosonic fields entering the path integral (\ref{S}). Here the conjugate variables $\zeta_{\gamma,j}$ and $\delta\rho_{\gamma,j}$ describe, respectively, the local phase and the density fluctuation of the bosonic condensates around the mean-field solution $\bar{\rho}=(g/2\Delta_{0})^{2}\sin^{2}(\bar{\theta})$. In the limit of $\delta\rho_{\gamma,j}/\bar{\rho}\ll 1$ one can expand the square root in Eq. (\ref{polar}) and keep only the quadratic terms of the density fluctuations. The latter condition can be fulfilled for large coupling $g\gg g_{\rm c}$ ($\bar{\rho}\gg 1$), where the quantum fluctuations are suppressed \cite{Porras2012}. Assuming that the spin and bosonic fields vary smoothly on the scale of lattice constant $a$ in a $d$ dimensional cubic lattice one can perform gradient expansion, such that in the symmetry broken phase $g>g_{{\rm c}}$ the continuum action becomes
\begin{eqnarray}
S&=&a^{-d}\int_{0}^{\beta}d\tau\int d^{d}x\{S_{\rm B}^{\prime}+\sum_{\gamma}\{{\rm i}\delta\rho_{\gamma}\frac{\partial\zeta_{\gamma}}{\partial\tau}+t\bar{\rho}a^{2}(\nabla\zeta_{\gamma})^{2}\notag\\
&&+\frac{ta^{2}}{4\bar{\rho}}(\nabla\delta\rho_{\gamma})^{2}+\sqrt{\bar{\rho}}\frac{g\sin(\bar{\theta})}{2}(\zeta_{\gamma}^{2}+\frac{\delta\rho_{\gamma}^{2}}
{4\bar{\rho}^{2}})\}+\Omega n_{+}^{\prime}n_{-}^{\prime}\notag\\
&&-\frac{g\cos(\bar{\theta})}{2\sqrt{\bar{\rho}}}n_{x}^{\prime}(\delta\rho_{{\rm r}}+\delta\rho_{{\rm l}})+g\sqrt{\bar{\rho}}n_{y}^{\prime}(\zeta_{{\rm r}}-\zeta_{{\rm l}})\}.\label{S_con}
\end{eqnarray}
Observe that the term $t\bar{\rho}a^{2}(\nabla\zeta_{\gamma})^{2}$ corresponds to the kinetic energy of a free particle with quadratic dispersion relation $\sim k^{2}$. Additionally, the spontaneous symmetry breaking gives rise to terms proportional to $\zeta_{\gamma}^{2}$ such that the system gain an energy gaps which are not vanish in the limit $k\rightarrow 0$. As a result of that the conjugate pairs $(\zeta_{{\rm r}},\delta\rho_{{\rm r}})$ and $(\zeta_{{\rm l}},\delta\rho_{{\rm l}})$ lead to two gapped amplitude modes in the spectra of cJT model. Indeed, the subsequent diagonalization of the action (\ref{S_con}) in the position-momentum representation gives an identical to $B_{a,a^{\prime}}^{(k)}$ matrix Eq. (\ref{B_matrix}) where the bosonic dispersion is replaced by its long-wave length limit, $\Delta_{k}\approx \Delta-td+t(ka)^{2}$.
\section{Measurement Protocol}\label{MP}
At the end we discuss the experimental verifiability of the presented results. For concreteness, we focus on the trapped ion-based realization of the cJT model. The protocol starts with the initialization of the linear ion crystal in the normal phase ground-state $g=0$, by laser cooling of the radial phonons to the motional ground-state and pumping spins to $\left|\downarrow_{j}\right\rangle$ state. After that the Jahn-Teller coupling is switched on and increases adiabatically to the desired regime $g>g_{\rm c}$. The quantum magnetic structural phase transition predicted by the mean-field solution (\ref{sp}) can be detected by measuring either the spin population or the phonon number. The antiferromagnetic spin order can be verified experimentally using a spin-dependent laser fluorescence, where the spins in the upper state emit light and appear bright, while the spins on the down state remain dark. The structural phase transition is related with the position reordering of the ion crystal into the zigzag configuration and creation of phonons in the lowest energy vibrational mode. These can be measured by laser induced fluorescence, which is imaged on a CCD camera and, respectively, with sideband spectroscopy which allows to determine the mean phonon number \cite{Haffner2008}. Finally, following the recent proposals \cite{Kurcz2014prl,Kurcz}, the dispersion relations of the excitations can be measured by the dynamical response of the system due to the weak coupling to the quantum probe.

\section{Conclusion}\label{C}

In conclusion, we have provided the path-integral formalism of the cJT model. Within the saddle-point approximation we have obtained the classical ground-state of the cJT model. The solution predicts a quantum magnetic structural phase transition with formation of ferromagnetic spin order and condensations of the two bosonic species in the lowest energy mode. We have calculated the elementary excitations of our model and found a linear gapless Goldstone mode and two gapped amplitude modes in the symmetry-broken phase.

\acknowledgments

This work has been supported by the EC Seventh Framework Programme under Grant Agreement No.270843 (iQIT).

\appendix
\section{Derivation of the energy spectrum}\label{A}
In order to determine the collective modes of the cJT model we define the harmonic oscillator degrees of freedom for each pair of conjugate variables using the relations
\begin{eqnarray}
&&\tilde{q}_{\epsilon,k}=\frac{1}{\sqrt{2\Delta_{k}}}(\delta\alpha_{\epsilon,-k}^{*}+\delta\alpha_{\epsilon,k}), \notag\\ &&\tilde{q}_{z,k}=\frac{1}{\sqrt{2\Omega}}(n_{+,-k}^{\prime}+n_{-,k}^{\prime}),
\end{eqnarray}
and respectively
\begin{eqnarray}
&&\tilde{p}_{\epsilon,k}={\rm i}\sqrt{\frac{\Delta_{k}}{2}}(\delta\alpha_{\epsilon,k}^{*}-\delta\alpha_{\epsilon,-k}), \notag\\ &&\tilde{p}_{z,k}={\rm i}\sqrt{\frac{\Omega}{2}}(n_{-,-k}^{\prime}-n_{+,k}^{\prime}).
\end{eqnarray}
These variables are used to express the action (\ref{S_k}) in the position-momentum representation which yield
\begin{eqnarray}
S&=&\int_{0}^{\beta}d\tau\sum_{k}\{\sum_{c}\{{\rm i}\frac{\partial \tilde{p}_{c,k}}{\partial\tau}\tilde{q}_{c,k}+\frac{1}{2}\tilde{p}_{c,k}\tilde{p}_{c,-k}\}+\frac{\Omega^{2}}{2}\tilde{q}_{z,k}\tilde{q}_{z,-k}\notag\\
&&+\frac{\Delta_{k}^{2}}{2}(\tilde{q}_{x,k}\tilde{q}_{x,-k}+\tilde{q}_{y,k}\tilde{q}_{y,-k})-2g_{{\rm c}}^{2}\sqrt{\frac{\Delta_{k}}{\Delta_{0}}}\tilde{q}_{x,k}\tilde{q}_{z,-k}\notag\\
&&-\sqrt{\frac{\Delta_{0}}{\Delta_{k}}}\tilde{p}_{y,-k}\tilde{p}_{z,k}.\label{S_pm}
\end{eqnarray}
The action (\ref{S_pm}) describes a collection of coupled oscillators. In order to decouple momentum dependent couplings between the different oscillators in (\ref{S_pm}) we perform transformation of the position variables
\begin{equation}
\left[\begin{array}{c}
\tilde{q}_{x,k}\\\tilde{q}_{y,k}\\\tilde{q}_{z,k}
\end{array}\right]=\frac{1}{\sqrt{2}}\left[
\begin{array}{ccc}
\sqrt{2} & 0 & 0\\
0 & -m_{+}(k)^{-1/2} & m_{-}(k)^{-1/2} \\
0 & m_{+}(k)^{-1/2} & m_{-}(k)^{-1/2}%
\end{array}%
\right]\left[\begin{array}{c}
q_{1,k}\\q_{2,k}\\q_{3,k}
\end{array}\right]\label{q}
\end{equation}%
and respectively of the momentum variables
\begin{equation}
\left[\begin{array}{c}
\tilde{p}_{x,k}\\\tilde{p}_{y,k}\\\tilde{p}_{z,k}
\end{array}\right]=\frac{1}{\sqrt{2}}\left[
\begin{array}{ccc}
\sqrt{2} & 0 & 0\\
0 & -m_{+}(k)^{1/2} & m_{-}(k)^{1/2} \\
0 & m_{+}(k)^{1/2} & m_{-}(k)^{1/2}%
\end{array}%
\right]\left[\begin{array}{c}
p_{1,k}\\p_{2,k}\\p_{3,k}
\end{array}\right].\label{p}
\end{equation}
Using Eqs. (\ref{S_pm}), (\ref{q}) and (\ref{p}) we arrive to Eq. (\ref{S_final}).

\end{document}